%% file: main.tex
\documentclass[sigconf,nonacm]{acmart}

\usepackage{array,booktabs,amsmath,graphicx,fancyvrb,tabularx,longtable}
\usepackage{subcaption}
\usepackage[tableposition=top]{caption}

\AtBeginDocument{%
  \setcopyright{rightsretained}%
  \copyrightyear{2026}%
  \acmYear{2026}%
}

\AtBeginDocument{%
  }

\begin{document}

\title{Looking Into the Past: Eye Movements Characterize Elements of Autobiographical Recall in Interviews with Holocaust Survivors}

\author{Emily Zhou}
\email{emilyzho@usc.edu}
\orcid{0009-0009-5004-6588}
\affiliation{
  \institution{University of Southern California}
  \city{Los Angeles}
  \state{CA}
  \country{USA}
}

\author{Marcus Ma}
\email{mjma@usc.edu}
\orcid{}
\affiliation{
  \institution{University of Southern California}
  \city{Los Angeles}
  \state{CA}
  \country{USA}
}

\author{Kleanthis Avramidis}
\email{avramidi@usc.edu}
\orcid{}
\affiliation{
  \institution{University of Southern California}
  \city{Los Angeles}
  \state{CA}
  \country{USA}
}
  
\author{Gabor Mihaly Toth}
\affiliation{%
  \institution{University of Luxembourg}
  \city{Esch-sur-Alzette}
  \country{Luxembourg}
}

\author{Shrikanth Narayanan}
\email{shri@usc.edu}
\orcid{}
\affiliation{
  \institution{University of Southern California}
  \city{Los Angeles}
  \state{CA}
  \country{USA}
}

\renewcommand{\shortauthors}{Zhou et al.}

\begin{abstract}
Eye movement and memory retrieval are deeply and bidirectionally intertwined, however existing literature is generally confined to controlled lab settings. We investigate the relationship between eye gaze and memory recall in free-form autobiographical recall, which comprises both autonoetic consciousness—the ability to mentally place oneself in the past or future—and various affective states. 
Using a large video corpus of semi-naturalistic interviews with Holocaust survivors (\textit{N}=806), we examine eye movements with respect to episodic, semantic, affective, and temporal dimensions of traumatic and highly emotional autobiographical recall.
We observe gaze patterns vary significantly across certain temporal contexts, most prominently in vertical eye movements.
We additionally train intra-subject sequence models to predict temporal context of sentences from segments of gaze features, and find that eye movements entirely preceding sentence onset are sufficient for prediction.
Our results corroborate prior findings in literature linking eye movements to memory in controlled and semi-structured settings, reinforcing the role of eye gaze in retrieving and constructing memories, especially in highly emotional and remote memory recall. 
\end{abstract}



\maketitle

\section{Introduction}

We remember with our eyes. Eye movements play a vital role in memory recall; they change depending on the type of recall~\citep{martarelli2017time, servais2022gaze}, our emotions~\citep{elhaj2017emotion}, and who we speak to during recall ~\citep{kleinke1986gaze}. For example, the relative time period of events (past or future-oriented) and direction of attention (internal or external) influences the gaze direction~\citep{servais2022gaze, martarelli2017time, barker2026eye_episodic}. In controlled studies, memories retrieved in a fixed gaze condition have been found to be less detailed and are retrieved more slowly than in a free-gaze condition~\citep{lenoble2019fixation_autobio}.

However, prior work has been limited to associations between eye movement and memory recall at short temporal scales with high-quality, fine-grained recordings in controlled settings, thereby leaving a gap on how eye movement and memory recall interact in more naturalistic and ecologically valid settings. In this work, we investigate this relationship in the \textit{Voices} dataset obtained from the USC Shoah Foundation's Visual History Archive, consisting of thousands of video testimonies from Holocaust survivors. To the best of the authors' knowledge, these interviews elicit remote memory recall and emotion processing on a scale and intensity stronger than any existing lab-controlled gaze analyses.

In our analysis, we find that high-level gaze patterns are strongly associated with both the type of memory recall and the corresponding life period, within World War II and the Holocaust, even when accounting for the elapsed time in the interview. In particular, vertical eye movement relative to the interviewer emerges as a significant discriminative signal. These eye gaze patterns can also predict the temporal characteristics of recalled memories, as demonstrated by experiments that shift the gaze feature window backward in time to infer temporal context. The contributions of this work can be summarized as follows:

\input{tables_and_figures/label_dist}

\begin{itemize}
    \item We study the associations between eye movements and various characteristics of memory recall in semi-naturalistic, video-recorded interviews with Holocaust survivors, with narrations spanning from childhood to the present day.
    \item We quantify the effect of recalling different time periods on eye gaze relative to the conversational partner in naturalistic autobiographical recall. 
    Individuals tend to divert their gaze away from the interviewer right after sentence onset, and this effect is particularly strong when recalling memories from the remote past, i.e., their childhood.
    \item We train deep learning models to predict the temporal context of recalled content using coarsely sampled eye gaze from naturalistic video recordings.
\end{itemize}

\section{Related Work}

\subsection{Memory and Autobiographical Recall}

Autobiographical memory plays a fundamental role in constructing self-identity and psychological well-being \citep{conway2000autobio} and involves both semantic and episodic components, terms introduced by \citeauthor{tulving1972episodic}~\citep{tulving1972episodic, tulving2002episodic}.
\textit{Episodic} memory describes the system in which the brain encodes and stores information about specific events and the spatiotemporal relationships between the associated components.
On the other hand, \textit{semantic} memory describes the memory for general knowledge and language that does not require one to recall experiences of distinctive events.

In our experiments, we retroactively label interview transcripts using the definitions outlined by the Autobiographical Interview protocol, a well-established procedure for quantifying episodic and semantic aspects of autobiographical memory~\citep{levine2002aging_autobio}. 
In the Autobiographical Interview, episodic details are referred to as \textit{internal}, referring to components of an event that occurred at a specific time and place.
Semantic information is referred to as \textit{external}, details describe non-episodic components, such as facts not specific to an event, general knowledge, and repeated statements. 
We use these definitions when labeling transcripts for memory recall.

\begin{figure*}[t]
    \centering
    \includegraphics[width=0.95\textwidth]{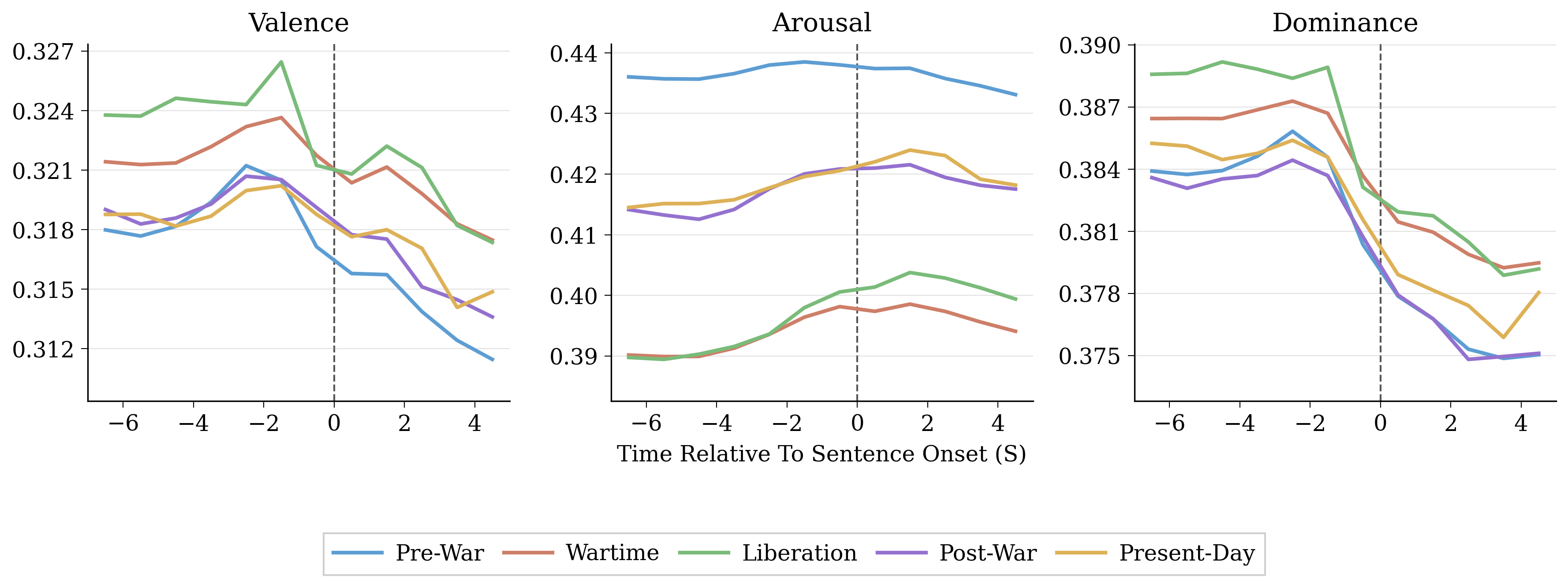} 
    \caption{Valence, Arousal, and Dominance estimated from audio recordings and aligned with transcripts.}
    \label{fig:affect}
\end{figure*}

\subsection{Gaze Patterns During Memory Recall}
Extensive studies on memory recall and eye gaze across controlled and semi-controlled conditions have found relationships between eye movement and various characteristics of memory recall. 
High-level features, such as gaze direction, are associated with temporal elements during both encoding and recall~\citep{martarelli2017time} as well as perceived direction of attention (external, \textit{i.e.}, objects, people, or internal, \textit{i.e.}, memories of personal experiences)~\citep{servais2022gaze}.
In a controlled encoding task of past versus future-oriented events about a fictitious character in a cohort of German speakers, \citeauthor{martarelli2017time}~\citep{martarelli2017time} found that eye gaze drifts to the right when recalling terms related to future events, but vertical eye movements did not display any significant differences.
Meanwhile, internal attention is strongly associated with an upward-averted gaze, while gaze remains around the horizontal axis when recalling external memories~\citep{servais2022gaze}. 

Fine-grained features, including fixations (temporary suspension of eye movement), saccades (involuntary, rapid eye movements between fixations), and blinks, are also associated with autobiographical recall.
In one study, individuals were asked to participate in an audio-guided tour of a museum and asked to freely recall details a week later~\citep{barker2026eye_episodic}. 
Mixed-effects models showed that saccades occur more frequently prior to episodic autobiographical details followed by a decrease prior to the next detail. 

Eye gaze has also been found to be a salient signal for detecting affect~\citep{soleymani2012, odwyer2017eye_speech_affect, elhaj2017emotion, Avramidis2026-eq, ma2026eye_ssl}.
When analyzing affective elements of memory recall, emotionally charged memories were associated with a higher frequency of fixations and saccades than neutral memories~\citep{elhaj2017emotion}.
Prior work using the \textit{Voices} data trained a self-supervised model to reconstruct eye movement, learning representations that successfully predicted speech-based affect~\cite{ma2026eye_ssl}.

\subsection{Gaze Patterns in Dyadic Interactions}
Dyadic interactions encompass a wide variety of settings, including turn-taking conversations, cooperative tasks, and interviews. 
In these settings, gaze direction is a critical social cue for establishing a turn in a conversation and gauging understanding of the interlocutor~\citep{jokinen2010turn}.
Under additional cognitive and social load, gaze aversion is particularly significant. 
When children were asked a series of increasingly difficult questions on arithmetic, language, and episodic and autobiographical memory~\citep{doherty2005gaze}, gaze aversion was the highest when autobiographical questions were asked (typically about recent classroom activities).
A study of gaze dynamics across two collaborative, conversation-based tasks in ~\citep{ho2015speaking} showed that speakers tended to begin their turn with as averted gaze, before turning their gaze back to their partner roughly 700ms into their turn. However, a gap in eye gaze literature exists at the intersection of memory recall and semi-naturalistic, dyadic interaction, namely how does conversational memory recall impact eye movements, when both memory recall and conversation have individually been shown to influence eye movements?

\section{Data Curation}
For our study, we select a subset of videos from the Shoah Foundation’s Visual History Archive, a collection of video testimonies with survivors of various genocides and persecution across the globe. We select the interviews of 978 Holocaust survivors (526 men, 452 women), where survivors recount their life experiences starting from early childhood, throughout World War II and the Holocaust, and up until the present day of the interview. In these testimonies, survivors recall traumatic experiences in internment camps, forced labor, and mass killings, as well as general life reflections. The interviews typically begin with a recall of the interviewee's childhood, which tend to be less temporally specific than memories from later in life. Topics near the end of the interview are more oriented around the present day (at the time of the interview), and are more philosophical, \textit{e.g.}, how they currently feel about their past, and what messages they may have to share.

Each video is about 30 minutes long and is recorded at 30 FPS and 320x240 pixel resolution. Each interviewee appears in multiple consecutive videos, totaling about two hours of video per interviewee. We also have transcripts annotated by human experts at the Shoah Foundation that are time-aligned on the word level.
After filtering (described in Section~\ref{sec:autobio_labels}), over 239,000 sentences related to autobiographical recall were used in our analysis.

\begin{figure*}[t]
    \centering
    \includegraphics[width=0.9\textwidth]{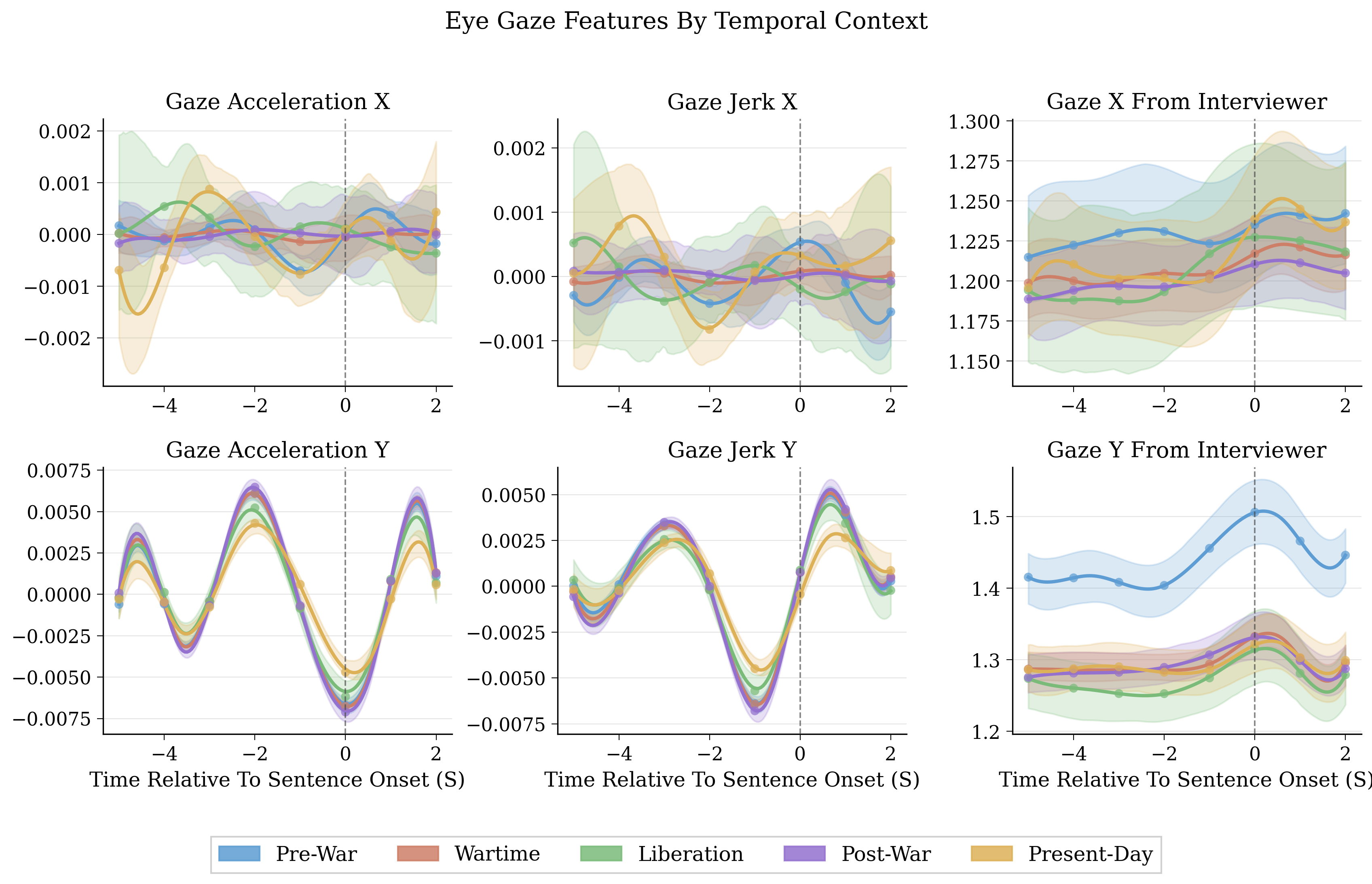} 
    \caption{Gaze dynamics (GAMM) across temporal contexts around sentence onset.}
    \label{fig:gamm}
\end{figure*}

\begin{figure*}[t]
    \centering
    \begin{subfigure}[b]{0.48\textwidth}
        \centering
        \includegraphics[width=\textwidth]{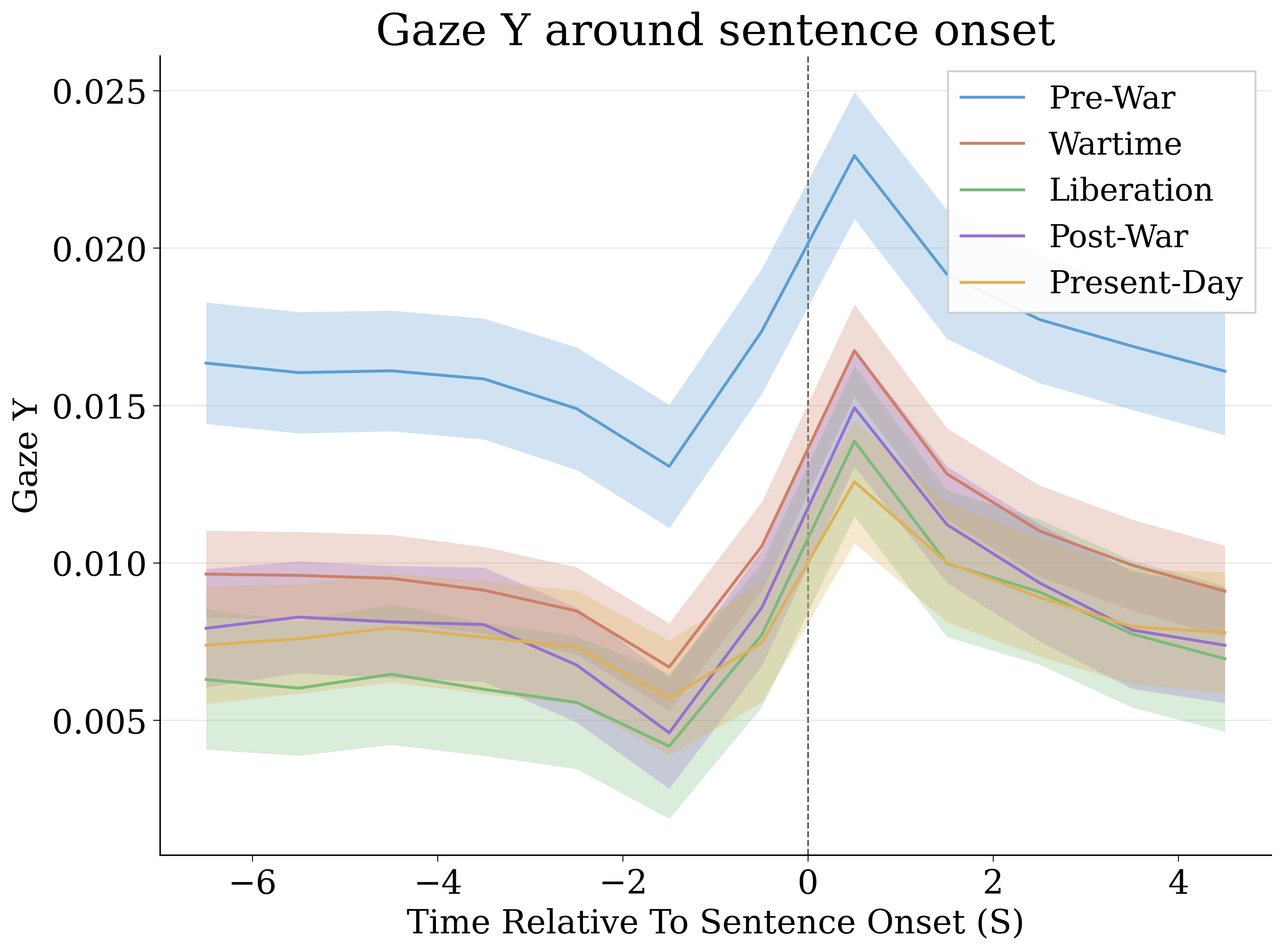}
        \caption{Vertical gaze direction in local coordinates.}
        \label{fig:first}
    \end{subfigure}
    \hfill 
    \begin{subfigure}[b]{0.48\textwidth}
        \centering
        \includegraphics[width=\textwidth]{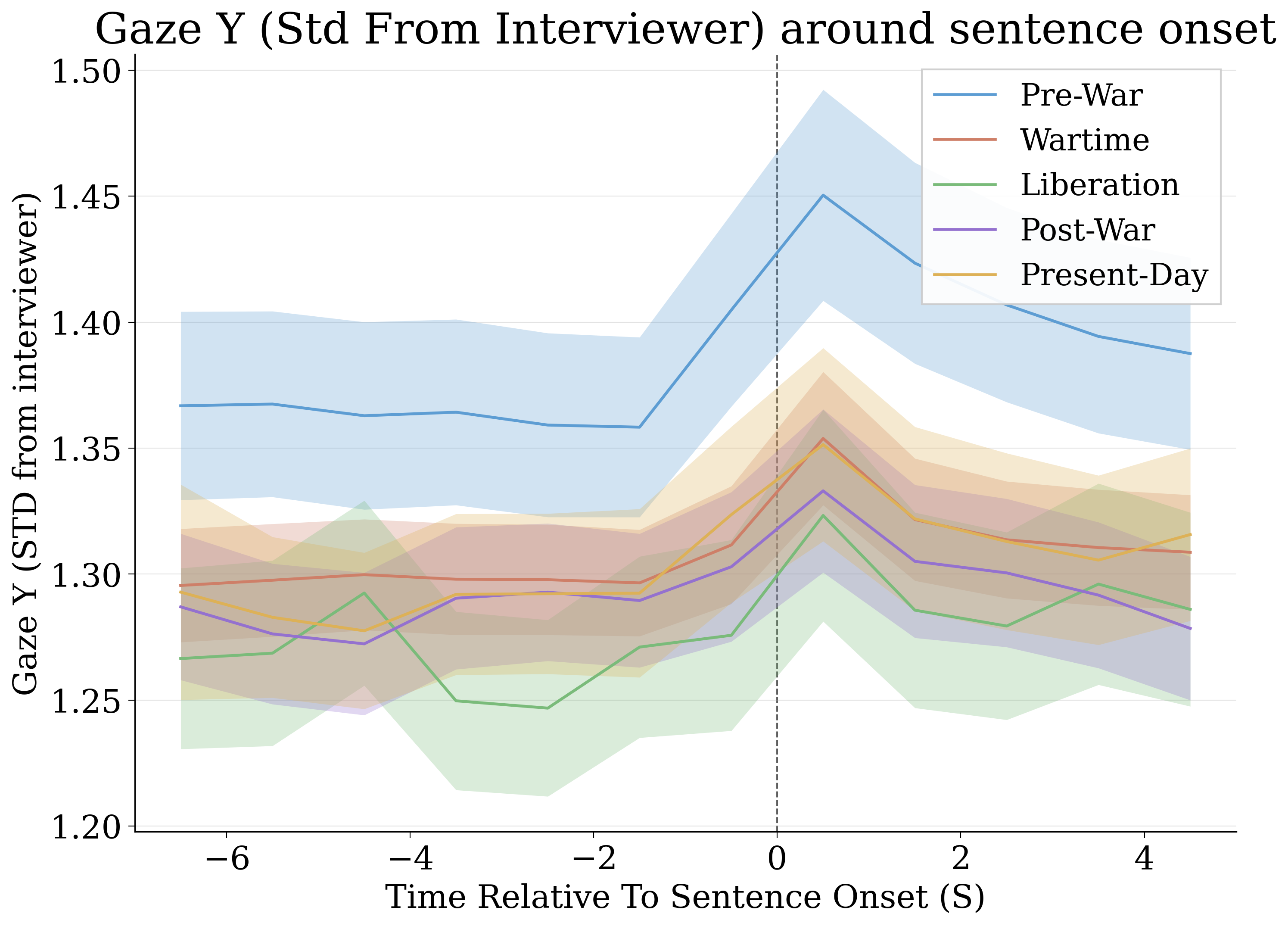}
        \caption{Vertical gaze distance from interviewer.}
        \label{fig:second}
    \end{subfigure}
    \caption{Temporal dynamics of vertical gaze relative to sentence onset differ significantly for the \textit{pre-war} context.}
    \label{fig:gaze_dynamics}
\end{figure*}

\subsection{Sentence-Level Annotations}
\subsubsection{Autobiographical Recall}
\label{sec:autobio_labels}
Recent studies have successfully adopted Large Language Models (LLMs) to automate the scoring of autobiographical recall~\citep{wardell2021ai_automated, van2024automated}.
In this work, we leverage the GPT-OSS-120B~\citep{gpt-oss-120b} model to extract characteristics of memory recall and temporal elements through prompting.
Sentences are labeled as either internal (episodic) or external (semantic) according to the definitions outlined in the Autobiographical Interview. 

Before we annotated all sentences, we performed iterative prompt engineering \cite{schulhoff2025promptreportsystematicsurvey} on a held-out validation set of 100 sentences. Two human annotators also labeled the same sentences. The final prompt, which included definitions and examples of internal/external memory recall sentences, achieved 81\% accuracy on human agreement with a Cohen's kappa of 0.62.

Each sentence is also labeled by the relevant time period in the interviewee's life, referred to as \textit{temporal context}: pre-war, wartime, liberation, post-war, present-day, and other. 
High-level measures of the richness of memory recall, duration and word count, are computed from the transcripts for each interviewee response.
To construct the final dataset, sentences shorter than 5~s or labeled as `other' were discarded due to their miscellaneous nature, and interviewees having fewer than the 10th quartile of sentences ($n{=}87$) were excluded.
This process resulted in a total of 806 interviewees and 239,353 sentences. 
The number of sentences per recall and temporal context type are shown in Table~\ref{tab:label_dist}.

\input{tables_and_figures/gaze_recall_type_stats}

\subsubsection{Affective Labels}
Labels of valance, arousal, and dominance (the VAD model)~\citep{mehrabian1996} are obtained from audio data~\citep{lertpetchpun2025}, time-aligned with the transcripts.
VAD measures are normalized to a range of [0, 1] and averaged across each sentence to obtain a proxy of felt emotion and emotional load during autobiographical recall.
Figure~\ref{fig:affect} visualizes changes in the three affective dimensions around sentence onset across the five temporal contexts.

\subsection{Eye Gaze Pre-processing}
\label{sec:eye_preprocess}
Eye and head position and angle were extracted using OpenFace 2.0~\citep{openface2}, resulting in frame-level vectors for gaze direction and head pose (X, Y, Z) in the same experimental setup described by \citeauthor{ma2026eye_ssl}~\cite{ma2026eye_ssl}.
Analysis of eye gaze typically uses saccade, fixation, and blink features, but the relatively lower quality and frame rate of this data prevents the accurate estimation of these features. 
Instead, we compute high-level features, localized with respect to the interviewee's head pose and interviewer's location. 

OpenFace returns eye gaze direction vectors in world coordinates. 
Since the recording camera is not set at the same position relative to the interviewee across videos, eye gaze direction vectors were first converted to a local coordinate system centered at the interviewee's face in each video. 
Segments with more than 30 consecutive frames of missing gaze data are discarded.
Next, the interviewer's location is estimated as the mode of eye gaze direction during segments of interviewer speech in the first ten minutes of each interview.
This value is then subtracted from each measurement to localize all data to the interviewer's position. 
Left and right eye gaze features are averaged and six features are calculated for each frame: 
\begin{itemize}
    \item (1--2) Gaze position in the X (horizontal) and Y (vertical) directions, represented in local coordinates.
    \item (3--6) Number of standard deviations (calculated per interviewee) away from the interviewer, separately for gaze position and angle in the X and Y directions. This encodes gaze information relative to the interviewer's position, while accounting for intra-subject variance.
\end{itemize}

Finally, the features are normalized via robust scaling.

\section{Experiments}
The relationship between eye gaze and memory has been studied extensively in controlled studies with short temporal scales~\citep{martarelli2017time, lenoble2019fixation_autobio, barker2026eye_episodic}, but not in more naturalistic, free-recall settings. 
To bridge this gap, we first conduct a series of statistical tests on the high-level eye gaze features to quantify the associations between the eye gaze features and recall type across different life periods. 
Next, end-to-end deep learning sequence models are trained to predict the temporal context of memory recall in autobiographical interviews from the eye gaze features, a five-class classification task. 

\begin{figure*}[t] %
    \centering
    \begin{subfigure}[b]{0.48\textwidth}
        \centering
        \includegraphics[width=\textwidth]{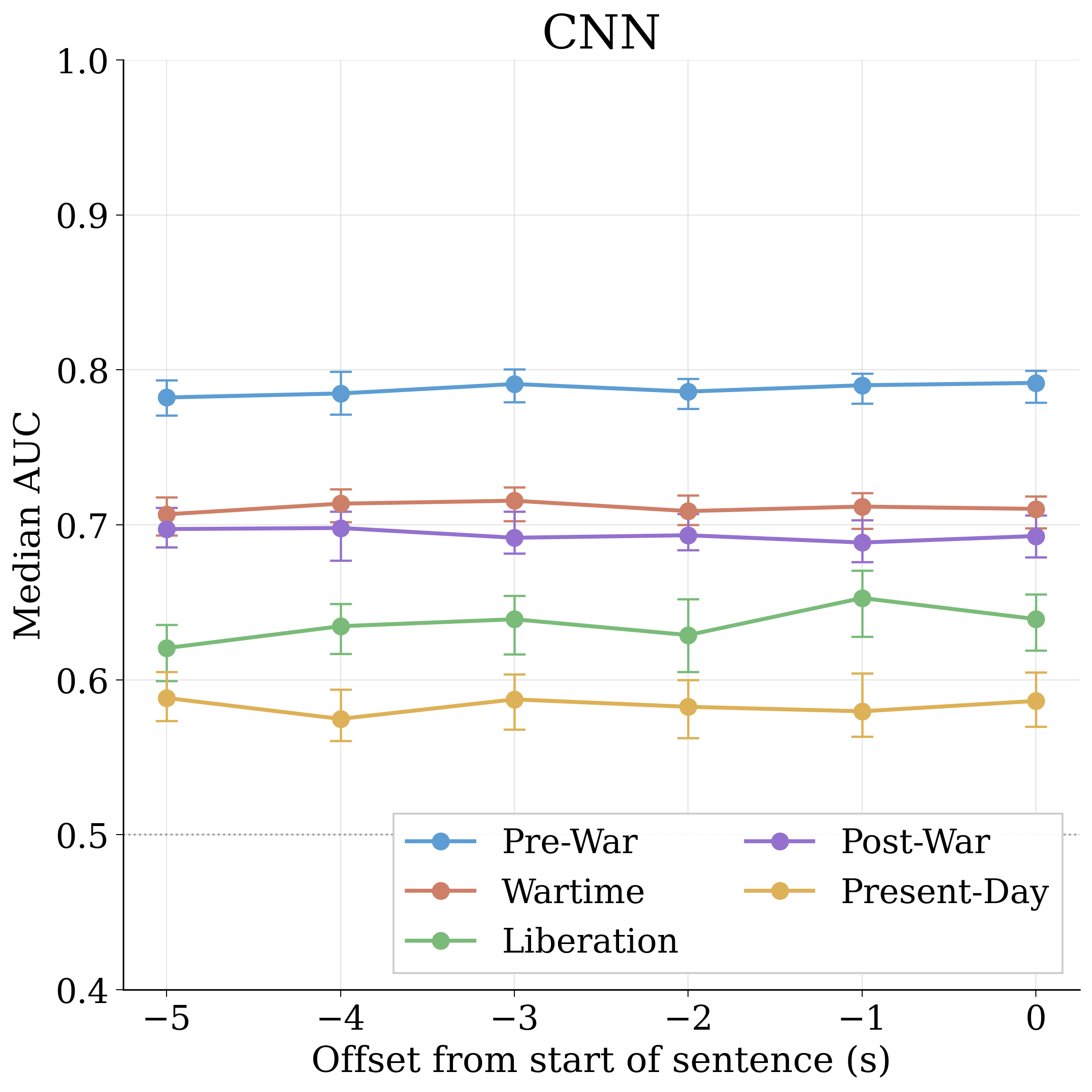}
        \caption{CNN performance across context window offsets.}
        \label{fig:first}
    \end{subfigure}
    \hfill 
    \begin{subfigure}[b]{0.48\textwidth}
        \centering
        \includegraphics[width=\textwidth]{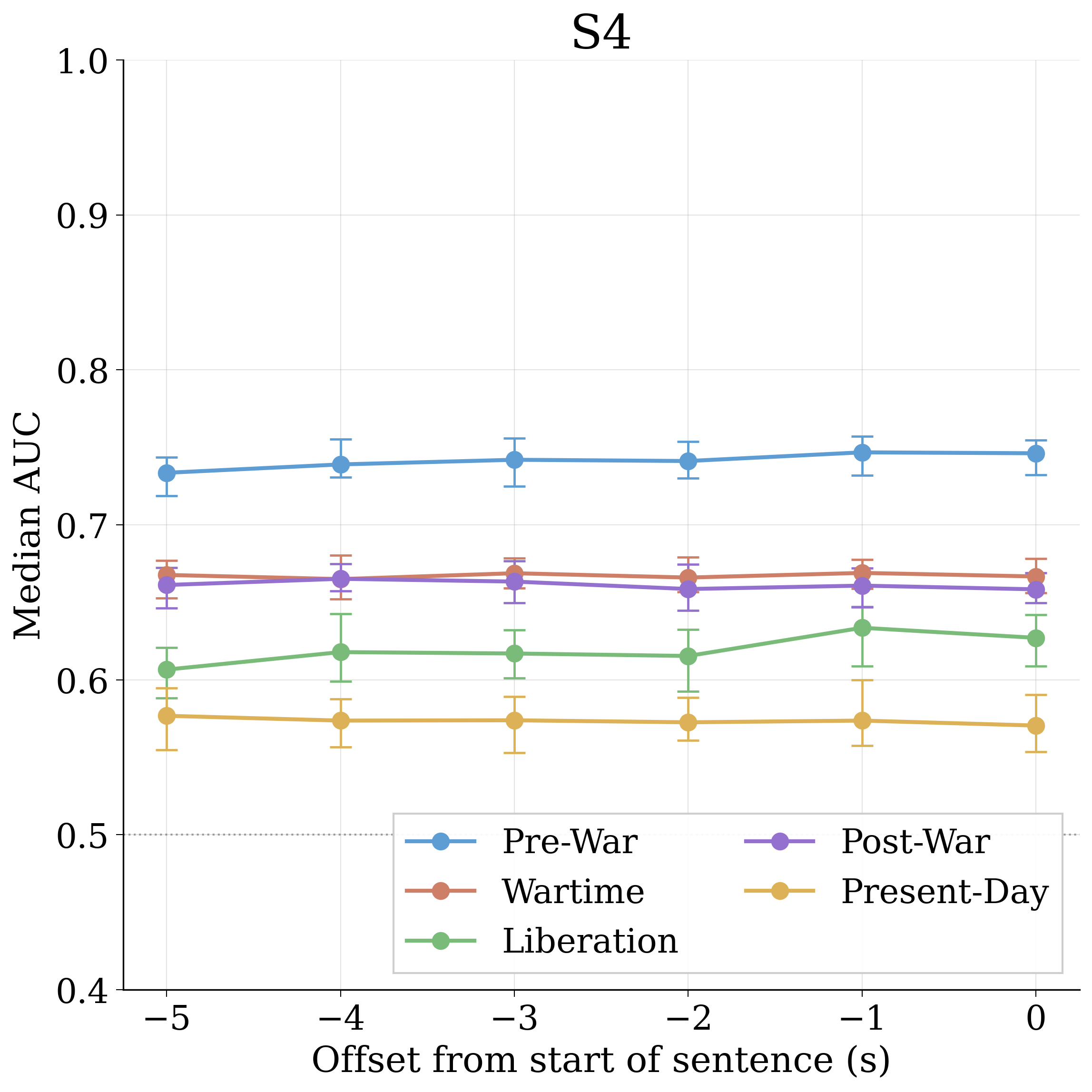}
        \caption{S4 performance across context window offsets.}
        \label{fig:second}
    \end{subfigure}
    \caption{Eye gaze preceding sentence onset predicts temporal context as effectively as eye gaze during speech. AUC (OvR) is reported, and baseline performance (0.5) is shown as a dashed line.}
    \label{fig:temp_context_class}
\end{figure*}

\subsection{Associating Gaze Features with Autobiographical Recall}
\subsubsection{Recall Type}
To quantify the association between eye movements and characteristics of autobiographical recall, Welch's t-test is performed to predict internal vs. external recall type from the gaze features. 
To evaluate differences in eye gaze that may be predictive of memory recall, measurements of each of the six continuous eye gaze features (see Section \ref{sec:eye_preprocess}) are averaged over the five seconds immediately preceding sentence onset. 

\subsubsection{Temporal Context}
To model the non-linear trajectories of the various eye gaze features, 
Generalized Additive Mixed Models (GAMMs) are fitted to model gaze features across the different temporal contexts. 
Additional features are calculated from the initial gaze features: velocity, acceleration, and jerk in the horizontal and vertical directions.
95\% confidence intervals are estimated via cluster bootstrapping with a sample size of 300 interviewees, effectively accounting for within-subject correlations. The resulting
GAMM trajectories are plotted in Figure~\ref{fig:gamm}. 

\subsection{End-to-End Prediction of Temporal Context}
To evaluate the predictive ability of eye gaze for memory recall, we train two deep learning models, a temporal CNN (TCNN)~\citep{tcnn} and a Structured State Space sequence (S4) model~\citep{gu2022s4}, to predict sentence-level temporal context using frame-level eye gaze features. For each temporal context (e.g., `wartime' and `post-war'), we predict a binary probability for that context occurring. We evaluate each temporal context individually in a One-vs-Rest binary setup and report AUC on this binary task.
Specifically, the 6 continuous eye gaze features over a $T_{context}$-second window are used to predict the recall type of each interviewee sentence.

The TCNN model consists of 6 layers, hidden dimension of 32, kernel size of 3, and dropout of 0.2, for a total of 16k trainable parameters.
The S4 model consists of 3 S4 blocks, hidden dimension of 32, and dropout of 0.2, with 32k trainable parameters.
Both models are trained using the AdamW optimizer with a learning rate of 1e-3 and a weight decay of 1e-2.
A separate model is trained for each interviewee in a 5-fold stratified cross validation setting across sentences with context length $T_{context}$ = 5 seconds.

The models are trained on a series of different context windows.
First, the start of the context window is aligned with the beginning of the target sentence for a pure classification task.
The window is shifted backwards with a one second time step, up to 7 seconds prior to sentence onset to obtain predictions from gaze windows preceding the sentence.

\input{tables_and_figures/ablation}

\subsection{Ablation of Horizontal and Vertical Axes}
Visual inspection of the vertical axis gaze dynamics (Figure~\ref{fig:gaze_dynamics}) show consistent increases in vertical direction, particularly away from the interviewer's location, around 1 second after sentence onset.
Based on this pattern, we further hypothesized that vertical gaze features alone in a 5-second window centered at 1 second after the start of a temporally relevant sentence is adequate for predicting temporal context.
To test this hypothesis, the same models are trained to predict temporal context using the $T_{context}$ = 5, centered at 1 second after sentence onset using only the three vertical or three horizontal gaze features.

\section{Results and Discussion}
Statistical analysis of the eye gaze features and their temporal dynamics around sentence onset are shown in Table~\ref{tab:gaze_recall_stats} and Figure~\ref{fig:gamm}, respectively. Temporal context prediction results over all classes with feature ablation are shown in Table~\ref{tab:ablation}, and performance metrics are visualized in Figures~\ref{fig:temp_context_class} and~\ref{fig:prior_context}.
Summary statistics are calculated across the full set by taking the mean across folds within interviewee before computing the median across all interviewees.

\subsection{Eye Gaze Patterns Differ Across Recall Type and Temporal Contexts}
Vertical eye gaze patterns differ significantly between internal and external recall, as well as across different temporal contexts. 
The Welch's t-test results (Table~\ref{tab:gaze_recall_stats}) show that both vertical gaze direction and vertical distance from interviewer are significantly higher for internal sentences, with much larger differences than horizontal gaze features. 
This aligns with prior findings of upward gaze aversion during memory recall~\citep{servais2022gaze}, especially when attention is directed internally, \textit{i.e.}, towards one's past experiences. 
We also did not observe significant changes in horizontal gaze direction.

Gaze trajectory plots (Figure~\ref{fig:gaze_dynamics}) show the same pattern in vertical gaze direction around sentence onset, with \textit{gaze\_y} and \textit{gaze\_y\_std} peaking about one second after the interviewee begins speaking. 
Both features exhibit similar patterns around sentence onset across all temporal contexts, but differ significantly when the individual is discussing the \textit{pre-war} time period, which extends from childhood up until World War II. 
The GAMM trajectories of acceleration, jerk, and standard deviations further support this, with highly consistent patterns and low variance in vertical features, but relatively higher spread in horizontal gaze features across the different contexts. 

\input{tables_and_figures/interview_progression_vs_temporal_context}

Importantly, interviewees tend to speak about their life periods in the same order, so the interview progression may be a potential confound. To test the effect of interview duration, we fit a linear mixed model to predict average vertical gaze position per sentence. The five temporal context categories and the sentence starting timestamp in seconds are the fixed effects, and the interviewee is treated as a random effect to control for inter-individual differences. We find that the progression of the interview (onset time) is a significant predictor, but with a minuscule effect ($\beta$ = -1.13e-6), and the likelihood ratio tests show that both the temporal context and elapsed time are significantly associated with gaze patterns. 

\subsection{Synchrony in Speech-Based Affect and Gaze}
The audio-based valence, arousal, and dominance trajectories around sentence onset, averaged across interviewees, were also visualized (Figure~\ref{fig:affect}). 
All contexts exhibited low valence (< 0.33), arousal (< 0.44), and dominance (< 0.39). 
Both valence and dominance typically drop within 2 seconds prior to the start of a sentence and tends to increase prior to the next sentence; the starting point of decrease aligns with the vertical gaze changes apparent in Figure~\ref{fig:gaze_dynamics}.
Arousal remains fairly constant, although there are clear differences in average arousal level across different temporal context groups, with arousal being highest when recalling \textit{pre-war} experiences and lowest when discussing \textit{wartime} and \textit{liberation}. 

\subsection{Gaze Aversion Increases at Sentence Onset}
The interviews constitute a form of dyadic conversation, in which eye gaze is known to both signal and respond to turn taking, but they are also dominated by the interviewee's autobiographical narrative as they recall their life experiences. 
Vertical gaze distance away from the interviewer's position begins to increase about 1.5 seconds prior to speaking and peaks just before one second after sentence onset, with a consistent pattern across all temporal contexts. 
The same patterns were seen during interviews of general questions of likes and dislikes~\citep{freeth2013affects} and cooperative gameplay~\citep{ho2015speaking}. 
This suggests that gaze aversion from a conversational partner extends beyond typical turn-taking and controlled interviews to autobiographical recall and narration about highly emotional and influential life events.

\begin{figure}[t]
    \centering
    \includegraphics[width=\columnwidth]{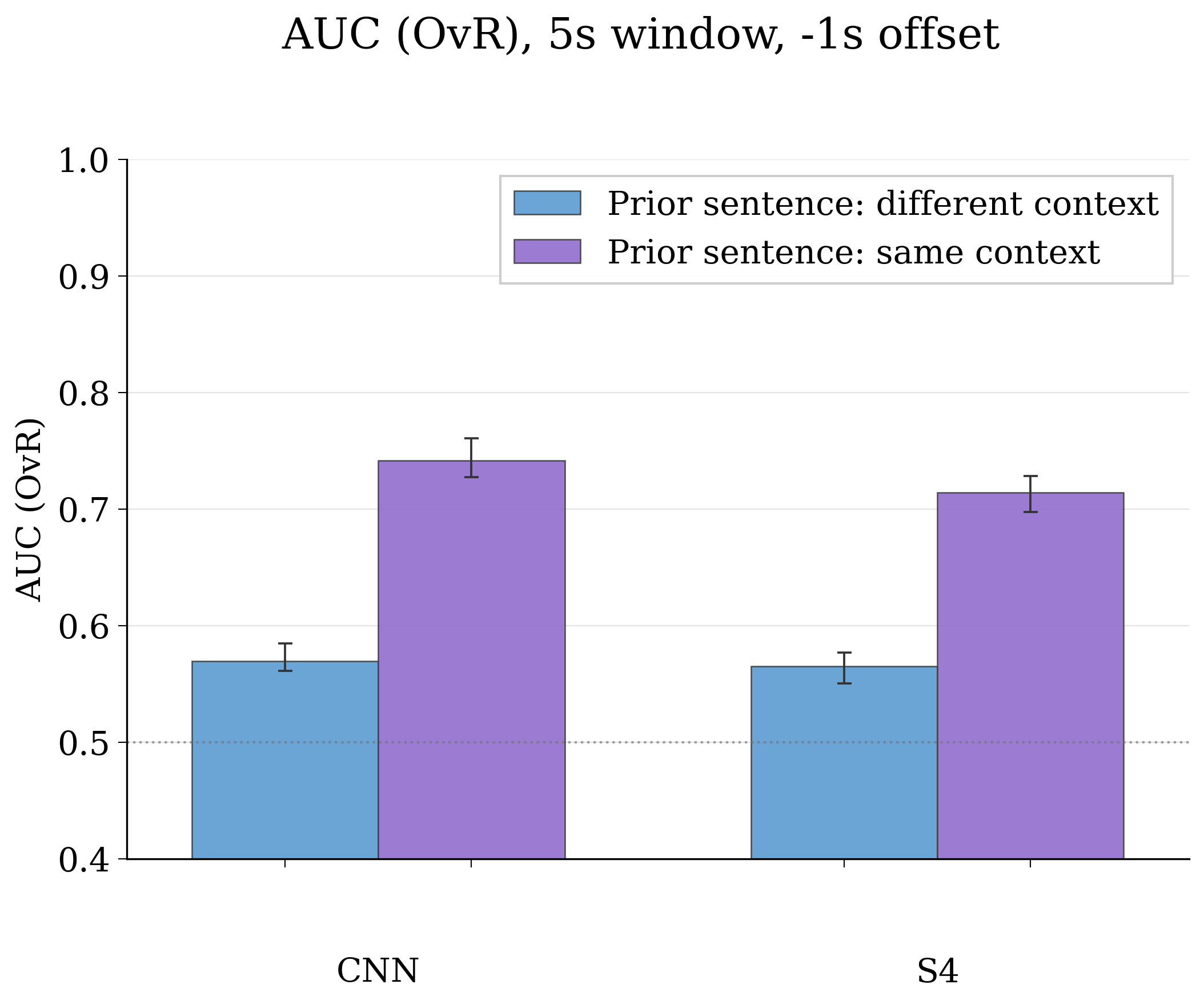} 
    \caption{Model prediction results by \textit{prior} context. Performance drops when temporal context changes, but remains above chance. '-1s offset' refers to the eye gaze window starting one second prior to sentence onset.}
    \label{fig:prior_context}
\end{figure}

\subsection{Eye Gaze Predicts Temporal Context of Autobiographical Recall}
We evaluate the predictive ability of eye gaze for temporal context of upcoming sentences by training TCNN and S4 sequence models on fixed-length, sliding windows starting prior to sentence onset. 
Results show that gaze windows entirely preceding the sentence are as predictive of temporal context as eye gaze during the sentence~\ref{fig:temp_context_class}.
Models achieve the highest performance for sentences related to the \textit{pre-war} time period, followed by similar performance on \textit{wartime} and \textit{post-war} (Figure~\ref{fig:temp_context_class}). 
It should be noted that class labels are imbalanced, with \textit{wartime} being discussed the most often; despite this, the models are able to predict the context of upcoming sentences well above chance performance. 
The \textit{present-day} context is more difficult to predict, with the lowest performance across all eye gaze settings.
This could be due to the more varied life experiences discussed, as well as the inclusion of various philosophical topics, thoughts, and feelings.

Additionally, we hypothesized that vertical gaze would be highly predictive of temporal context. 
The ablation study proves this hypothesis false, with the horizontal feature subset performing just as well as the vertical features.
The full feature set performed better than both subsets, demonstrating the importance of both vertical and horizontal gaze dynamics in differentiating between recall of different autobiographical time periods, even when statistical analyses does not show significant associations across all variables.

These results should be interpreted with the consideration of the duration of the autobiographical narratives. 
In particular, interviewees tend to speak continuously about the same time period for several minutes at a time, a possible explanation for the long predictive window.
When sentences with the same temporal context as the immediate prior sentence are filtered out, prediction AUC drops significantly (Table~\ref{fig:prior_context}).

\section{Limitations and Future Work}
Memory retrieval and recall is a highly complex process, involving neural mechanisms that relate prior knowledge, self-perception, temporal and spatial information, and affect.
Therefore, the relationship between eye gaze and memory, especially autobiographical recall, has largely been studied in controlled environments with structured recall tasks.
This work introduces an initial step towards an understanding of eye gaze features and autobiographical recall in a naturalistic, noisy setting. 
Labels generated by the LLM achieved high agreement with human annotators, but are not guaranteed to be correct, contributing additional noise.

In processing the eye gaze data, gaze features were normalized with respect to the interviewer's \textit{assumed} position, since only the interviewee is in frame. Interviewer position is estimated by taking the mode of the interviewee's gaze direction while the interviewer is speaking, but this cannot be directly verified.
As a result, noise may have contributed higher variance in the processed features.
The standard deviation in model performance across individuals also suggests high inter-individual differences in eye gaze patterns.

While this work focused on sentence-level analysis, examining eye gaze across full episodes of memory recall would provide additional insights into the complex relationships between the two, especially given the observed temporal consistency of the narratives in our dataset.
Future work could also explore multimodal sentiment analysis, including interactions between text, speech, eye gaze, and facial expressions across different affective states, and incongruence in displayed emotion across these modalities.

\section{Conclusion}
This work, to the best of the authors' knowledge, contributes the first study of eye gaze and characteristics of memory recall in autobiographical recall at this scale.
Our experiments elucidate the relationships between eye movements and temporal characteristics of autobiographical recall and demonstrate the predictive ability of gaze patterns across temporal contexts. 
Vertical gaze features differ between sentences of internal versus external attention, and different time periods throughout life. 
High-level gaze features are also predictive of the temporal context of recalled experiences with respect to the Holocaust and World War II. 
Despite the noisy setting, the reported results align with established findings in literature on the close relationship between eye gaze and memory.

\section{Safe and Responsible Innovation}
The videos used in the analyses were provided by the USC Shoah Foundation as part of the Visual History Archive\footnote{\url{https://vha.usc.edu/home}}, and no personally identifiable information is made available. 
This work focuses on the retelling of life events from survivors of the Holocaust, a tragic event in human history. 
Any analysis on these 
data, including the findings presented in this work, must be interpreted with careful ethical consideration and are limited to the temporal contexts defined by the historical events that shaped these individuals' lives. 

\bibliographystyle{ACM-Reference-Format}
\bibliography{references}

\end{document}

%% file: tables_and_figures/label_dist.tex
\begin{table*}[ht]
\centering
\caption{Distribution of Recall Type and Temporal Contexts}
\label{tab:label_dist}
\begin{tabular}{@{}l r r r r r r@{}}
\toprule
& \multicolumn{5}{c}{\textbf{Temporal Context}} & \\ \cmidrule(lr){2-6}
\textbf{Recall Type} & \textbf{Pre-war} & \textbf{Wartime} & \textbf{Liberation} & \textbf{Post-war} & \textbf{Present} & \textbf{Total} \\ \midrule
Internal & 12,305 & 105,602 & 5,308 & 21,098 & 1,311 & 145,624 \\
External & 25,852 & 51,276  & 919   & 8,506  & 7,176 & 93,729  \\ \midrule
\textbf{Total} & 38,157 & 156,878 & 6,227 & 29,604 & 8,487 & 239,353 \\ \bottomrule
\end{tabular}
\end{table*}

%% file: tables_and_figures/gaze_recall_type_stats.tex
\begin{table}[ht]
\centering
\caption{Welch's t-test Results for Eye Gaze Features Predicting Internal vs. External Recall. Features are averaged over 5 seconds immediately preceding sentence onset.}
\label{tab:gaze_recall_stats}
\small
\begin{tabular}{@{}l r r r@{}}
\toprule
\textbf{Feature} & \textbf{$t$} & \textbf{$p$-value} & \textbf{Mean (SD)} \\ \midrule
Gaze X           & -2.51 & 0.0122 & 0.0022 (0.043) \\
Gaze Y           & 18.75 & < \textbf{0.001} & 0.0085 (0.033) \\
Gaze X SD        & -0.92 & 0.3586 & 1.2022 (0.651) \\
Gaze Y SD        & 10.23 & < \textbf{0.001} & 1.3078 (0.627) \\
Gaze Angle X SD  & -2.34 & 0.0193 & 0.9932 (1.030) \\
Gaze Angle Y SD  & 5.08  & < \textbf{0.001} & 0.8326 (0.926) \\ \bottomrule
\addlinespace
\multicolumn{4}{l}{\footnotesize \textit{Note:} \textit{SD} refers to standard deviations away from the interviewer.}
\end{tabular}
\end{table}

%% file: tables_and_figures/ablation.tex
\begin{table}[t]
\centering
\caption{Ablation on Gaze Dimension for Temporal Context Prediction. Input is a 5-second window centered at 1 second after sentence onset. Best performance is denoted in bold.}
\label{tab:ablation}
\small

\begin{tabular}{@{}lccc@{}}
\toprule
\textbf{Input Features (\#)} & \textbf{ACC} & \textbf{Weighted F1} & \textbf{AUC (OvR)} \\
\midrule
Baseline & 0.6750 & 0.5440 & 0.5000 \\
\midrule
Horizontal (3) & 0.7082 (0.1301) & 0.6410 (0.1435) & 0.6642 (0.0937) \\
Vertical (3) & 0.7137 (0.1286) & 0.6406 (0.1441) & 0.6679 (0.0953) \\
Both (6) & \textbf{0.7171} (0.1266) & \textbf{0.6620} (0.1386) & \textbf{0.6969} (0.0959) \\
\bottomrule
\end{tabular}

\vspace{2pt}
\parbox{\linewidth}{\footnotesize \textit{Note:} The baseline condition represents a naive prediction based on class frequencies. All results reported as median across subjects (standard deviation).}
\end{table}

%% file: tables_and_figures/interview_progression_vs_temporal_context.tex
\begin{table}[h!]
\centering
\caption{Linear Mixed-Effects Model for Mean Vertical Gaze as a function of temporal context (pre-war baseline) and sentence timing  ($N = 245,370$ observations)}
\label{tab:int_duration_vs_temp_context}
\begin{tabular}{@{} l r c @{}}
\toprule
\textbf{Predictor} & \textbf{Coefficient $\beta$} & \textbf{Std. Error} \\
\midrule
\textit{Fixed Effects} & & \\
Intercept (Pre-war)*** & 0.01929 & 0.00073 \\
Wartime*** & -0.00580 & 0.00017 \\
Liberation*** & -0.00448 & 0.00038 \\
Post-war*** & -0.00516 & 0.00024 \\
Present-day*** & -0.00547 & 0.00033 \\
Sentence onset (s)*** & -1.13e-06 & 2.54e-08 \\
\midrule
\textit{Random Effects} & & \\
Interviewee ID ($\sigma^2_u$) & 0.000458 & \\
Residual ($\sigma^2_\epsilon$) & 0.000695 & \\
\midrule
\textit{Model Fit} & & \\
ICC & 0.3973 & \\
LRT: Content Domain*** & $\chi^2(4) = 1188.7$ & \\
LRT: Interview Onset*** & $\chi^2(1) = 1952.5$ & \\
\bottomrule
\end{tabular}

\vspace{2pt}
\parbox{\linewidth}{\footnotesize \hspace{0.25cm}\textit{Note:} *** $p < 0.001$. Baseline for temporal context dummy coding is 'pre-war'.} \\
\end{table}